\documentclass[slac_one]{revtex4}
\usepackage{graphicx}
\usepackage{fancyhdr}
\newcommand{\befig}{\begin{figure}}
\newcommand{\efig}{\end{figure}}
\newcommand{\betab}{\begin{table}}
\newcommand{\etab}{\end{table}}
\newcommand{\barray}{\begin{array}}
\newcommand{\earray}{\end{array}}
\newcommand{\be}{\begin{equation}}
\newcommand{\ee}{\end{equation}}
\newcommand{\bea}{\begin{eqnarray}}
\newcommand{\eea}{\end{eqnarray}}
\newcommand{\benn}{\begin{displaymath}}
\newcommand{\eenn}{\end{displaymath}}
\newcommand{\beann}{\begin{eqnarray*}}
\newcommand{\eeann}{\end{eqnarray*}}
\newcommand{\gtsim}{\gtrsim}

\newcommand{\Order}{{\cal O}}   
\newcommand{\keV}{\mbox{keV}}
\newcommand{\MeV}{\mbox{MeV}}
\newcommand{\GeV}{\mbox{GeV}}

\newcommand{\MPl}{\mathrm{M}_{\mathrm{Pl}}}

\newcommand{\gl}{\ensuremath{\tilde{g}}}
\newcommand{\ax}{\ensuremath{\tilde{a}}}
\newcommand{\gr}{\ensuremath{\tilde{G}}}
\newcommand{\st}{\ensuremath{\tilde{\tau}}}
\newcommand{\Bi}{\ensuremath{\tilde{B}}}

\newcommand{\stau}{{\widetilde \tau}}
\newcommand{\gravitino}{{\widetilde{G}}}
\newcommand{\axino}{{\widetilde a}}

\newcommand{\mgravitino}{m_{\widetilde{G}}}

\begin{document}

\title{{\small{2005 International Linear Collider Workshop - Stanford,
U.S.A.}}\\ 
\vspace{12pt}
Collider Signatures of Axino and Gravitino Dark Matter} 

\author{Frank Daniel Steffen}
\affiliation{DESY Theory Group, Notkestrasse 85, 22603 Hamburg, Germany}

\begin{abstract}
  
  The axino and the gravitino are extremely weakly interacting
  candidates for the lightest supersymmetric particle (LSP).  We
  demonstrate that either of them could provide the right amount of
  cold dark matter. Assuming that a charged slepton is the
  next-to-lightest supersymmetric particle (NLSP), we discuss how NLSP
  decays into the axino/gravitino LSP can provide evidence for
  axino/gravitino dark matter at future colliders. We show that these
  NLSP decays will allow us to estimate the value of the Peccei--Quinn
  scale and the axino mass if the axino is the LSP.  In the case of
  the gravitino LSP, we illustrate that the gravitino mass can be
  determined. This is crucial for insights into the mechanism of
  supersymmetry breaking and can lead to a microscopic measurement of
  the Planck scale.

\end{abstract}

\maketitle

\thispagestyle{fancy}

\section{INTRODUCTION}

A key problem in cosmology is the understanding of the nature of cold
dark matter. In supersymmetric extensions of the Standard Model, the
lightest supersymmetric particle (LSP) is stable if $R$-parity is
conserved~\cite{Nilles:1983ge+X}. An electrically and color neutral
LSP thus appears as a compelling solution to the dark matter problem.
The lightest neutralino is such an LSP candidate from the minimal
supersymmetric standard model (MSSM). Here we consider two
well-motivated alternative LSP candidates beyond the MSSM: the axino
and the gravitino.

In the following we introduce the axino and the gravitino.  We review
that axinos/gravitinos from thermal production in the early Universe
can provide the right amount of cold dark matter depending on the
value of the reheating temperature after inflation and the
axino/gravitino mass. For scenarios in which a charged slepton is the
next-to-lightest supersymmetric particle (NLSP), we discuss signatures
of axinos and gravitinos at future colliders.

\section{AXINOS AND GRAVITINOS}

The axino $\axino$~\cite{Nilles:1981py+X,Tamvakis:1982mw,Kim:1983ia}
appears (as the spin-1/2 superpartner of the axion) once the MSSM is
extended with the Peccei--Quinn mechanism~\cite{Peccei:1977hh+X} in
order to solve the strong CP problem. Depending on the model and the
supersymmetry (SUSY) breaking scheme, the axino mass~$m_{\axino}$ can
range between the eV and the GeV
scale~\cite{Tamvakis:1982mw,Nieves:1985fq+X,Rajagopal:1990yx,Goto:1991gq+X}.
The axino is a singlet with respect to the gauge groups of the
Standard Model. It interacts extremely weakly as its interactions are
suppressed by the Peccei--Quinn
scale~\cite{Sikivie:1999sy,Eidelman:2004wy} $f_a\gtrsim 5\times
10^9\,\GeV$.  The detailed form of the axino interactions depends on
the axion model under consideration.  We focus on hadronic, or KSVZ,
axion models~\cite{Kim:1979if+X} in a SUSY setting, in which the axino
couples to the MSSM particles only indirectly through loops of
additional heavy KSVZ (s)quarks.

The gravitino $\gravitino$ appears (as the spin-3/2 superpartner of
the graviton) once SUSY is promoted from a global to a local symmetry
leading to supergravity (SUGRA)~\cite{Wess:1992cp}. The gravitino
mass~$m_{\gravitino}$ depends strongly on the SUSY-breaking scheme and
can range from the eV scale to scales beyond the TeV
region~\cite{Nilles:1983ge+X,Dine:1994vc+X,Randall:1998uk+X,Buchmueller:2005rt}.
In particular, in gauge-mediated SUSY breaking
schemes~\cite{Dine:1994vc+X}, the mass of the gravitino is typically
less than 100~MeV, while in gravity-mediated
schemes~\cite{Nilles:1983ge+X} it is expected to be in the GeV to TeV
range. Also, the gravitino is a singlet with respect to the gauge
groups of the Standard Model. Its interactions---given by the SUGRA
Lagrangian---are suppressed by the (reduced) Planck
scale~\cite{Eidelman:2004wy} $\MPl=2.4\times 10^{18}\,\GeV$. Once SUSY
is broken, the extremely weak gravitino interactions are enhanced (for
small values of the gravitino mass) through the super-Higgs mechanism.

\section{AXINOS AND GRAVITINOS AS COLD DARK MATTER}

Because of their extremely weak interactions, the temperature $T_D$ at
which axinos/gravitinos decouple from the thermal plasma in the early
Universe is very high. For example, an axino decoupling temperature of
$T_D \approx 10^9\,\GeV$ is obtained for $f_a = 10^{11}\,\GeV$.
Gravitinos with $m_{\gr} \gtsim 10\,\MeV$ decouple at similar or even
higher temperatures. 
Below the decoupling temperature, axinos/gravitinos can be produced in
thermal reactions in the hot MSSM plasma. The thermal axino/gravitino
production rate at high temperatures can be computed in a
gauge-invariant way with the Braaten--Yuan prescription and hard
thermal loop resummation, which takes into account Debye screening in
the plasma.  Assuming that inflation has diluted away any primordial
axino/gravitino abundance, axino/gravitino-disappearance processes are
negligible for a reheating temperature $T_R$ sufficiently below $T_D$.
The corresponding Boltzmann equation can be solved analytically. This
leads to the following results for the relic densities
$\Omega_{\ax/\gr}h^2$ of stable LSP axinos/gravitinos from thermal
production~\cite{Bolz:2000fu,Brandenburg:2004du+X}
\bea\label{Eq:Omegah2_Axino}
        \Omega_{\ax}h^2
        & = &  
        5.5\,g^6 \ln\left( \frac{1.108}{g}\right) 
        \bigg(\frac{m_{\ax}}{0.1~\GeV}\bigg)\!
        \left(\frac{10^{11}\,\GeV}{f_a/N}\right)^{\! 2}\!\!
        \left(\frac{T_R}{10^4\,\GeV}\right) \, ,
\\ \label{Eq:Omegah2_Gravitino}
        \Omega_{\gr}h^2 
        & = & 
        0.12 \, g^2 \ln\!\left( \frac{1.163}{g}\right) 
        \left( 1 + {m^2_{\gl} \over 3 \mgravitino^2} \right)
        \left({\mgravitino \over 100~\GeV}\right)
        \left(\frac{T_R}{10^{10}\,\GeV}\right) \, ,
\eea
where $h$ parametrizes the Hubble constant $H_0 =
100\,h\,\mbox{km/s/Mpc}$, $g$ is the strong coupling, $N$ the number
of heavy KSVZ (s)quark flavors, and $m_{\gl}$ the gluino mass.  Note
that $\Omega_{\gr}h^2$ increases with decreasing $\mgravitino$ due to
enhanced gravitino interactions.  The axino couplings do not depend on
$m_{\ax}$ so that $\Omega_{\ax}h^2$ increases with increasing
$m_{\ax}$.

\begin{figure*}
\includegraphics[width=6.7cm]{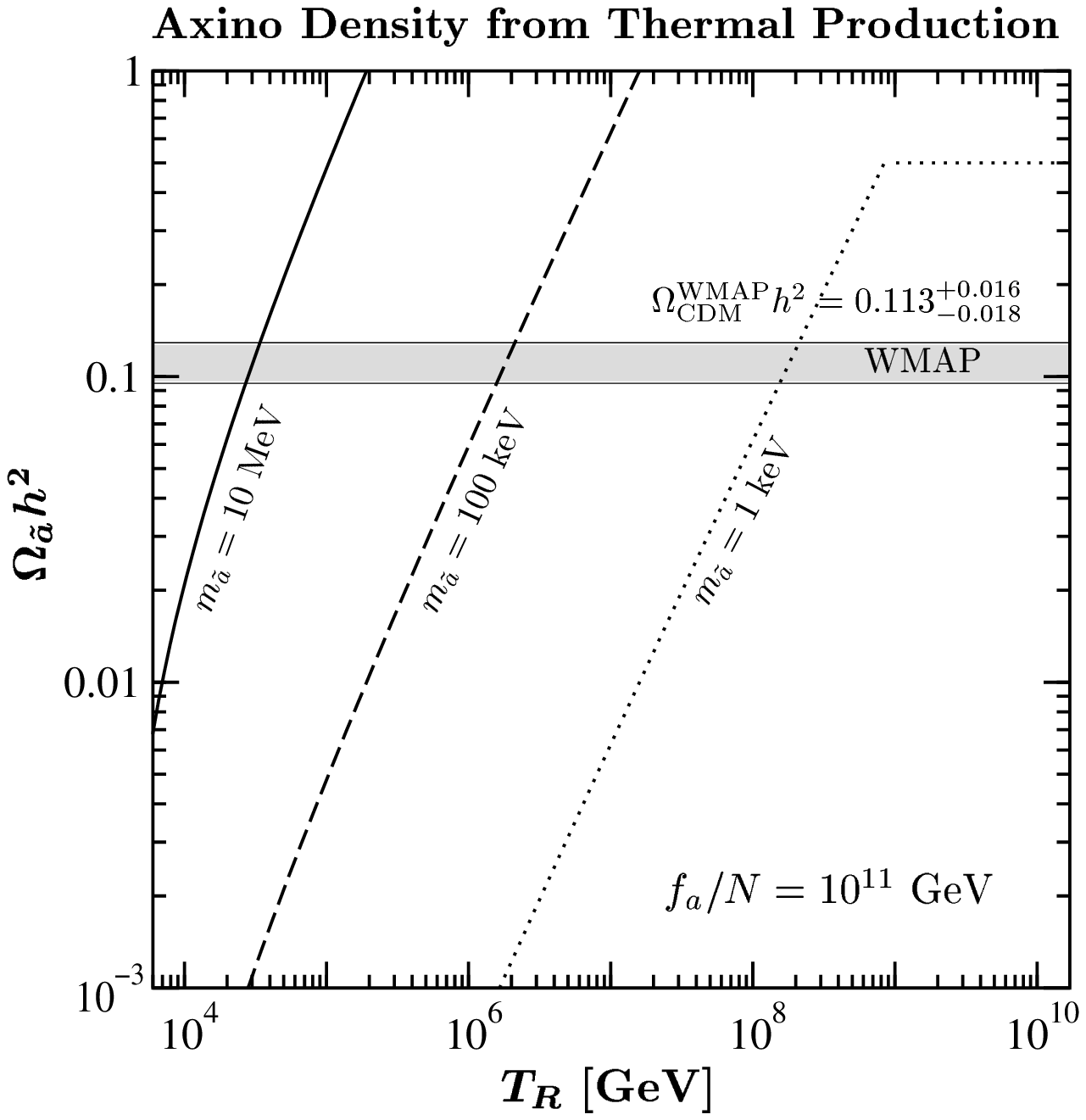}
\hskip 1.5cm
\includegraphics[width=6.7cm]{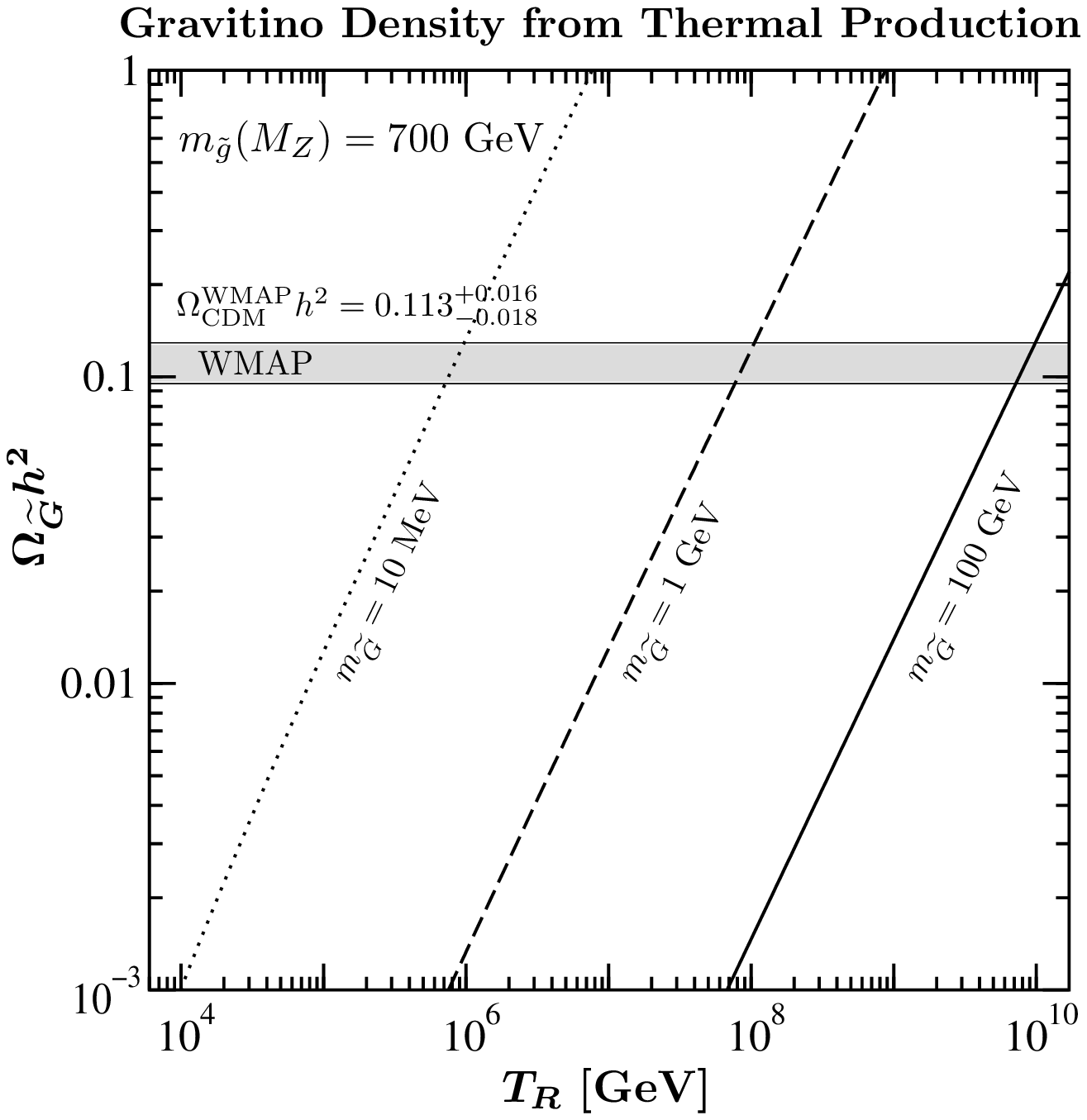}
\caption{
  Relic densities of axino LSPs (left) and gravitino LSPs (right) from
  thermal production in the early Universe.\label{Fig:DM_from_TP}}
\end{figure*}

In Fig.~\ref{Fig:DM_from_TP} we illustrate $\Omega_{\ax/\gr}h^2$ as a
function of $T_R$ for different values of $m_{\ax/\gr}$, where $f_a/N
= 10^{11}\,\GeV$ and $m_{\gl}(M_Z)=700\,\GeV$.
The running of the strong coupling and the gluino mass is taken into
account by replacing $g$ and $m_{\gl}$
in~(\ref{Eq:Omegah2_Axino},\ref{Eq:Omegah2_Gravitino}) respectively
with $g(T_R)=[g^{-2}(M_Z)\!+\!3\ln(T_R/M_Z)/(8\pi^2)]^{-1/2}$ and
$m_{\gl}(T_R)=[g(T_R)/g(M_Z)]^2 m_{\gl}(M_Z)$, where $M_Z$ is the
Z-boson mass and $g^2(M_Z)/(4\pi)=0.118$.
%
%
For $T_R$ above $T_D \approx 10^9\,\GeV$, $\Omega_{\ax}h^2$ is given
by the equilibrium number density of a relativistic Majorana fermion
and thus independent of $T_R$ as shown for $m_{\ax} = 1\,\keV$. There
will be a smooth transition instead of a kink once the
axino-disappearance processes are taken into account.  The grey band
indicates the WMAP result\,\cite{Spergel:2003cb} on the cold dark
matter density (2$\sigma$ error)
$\Omega_{\mathrm{CDM}}^{\mathrm{WMAP}} h^2 = 0.113^{+0.016}_{-0.018}$.
Axinos give the right amount of cold dark matter for
$(m_{\ax},T_R)=(100\,\keV,\,10^6\,\GeV)$.  Higher values of $T_R$ are
problematic as $m_{\ax}$ becomes too light to explain structure
formation. In contrast, gravitinos could provide the right amount of
cold dark matter for combinations with a very high reheating
temperature such as $(m_{\gr},T_R)=(100\,\GeV,\,10^{10}\,\GeV)$.
Nevertheless, other gravitino cold dark matter scenarios---such as
$(m_{\gr},T_R)=(10\,\MeV,\,10^6\,\GeV)$---are also viable.

\section{COLLIDER SIGNATURES OF AXINOS AND GRAVITINOS}

As a consequence of the extremely weak couplings, the direct
production of axino/gravitino LSPs at colliders is strongly
suppressed. Furthermore, the NLSP typically has a long lifetime, which
is subject to cosmological constraints as discussed, for example,
in~\cite{Covi:2004rb,Feng:2004mt}. At future colliders one expects a
large sample of such quasi-stable NLSPs if the NLSP belongs to the
MSSM spectrum. Each NLSP will eventually decay into the
axino/gravitino LSP. These decays can provide signatures of
axinos/gravitinos and other insights into physics beyond the MSSM
(cf.~\cite{Stump:1996wd,
  Ambrosanio:2000ik,Buchmuller:2004rq,Brandenburg:2005he} and
references therein). We concentrate on results extracted from
Ref.~\cite{Brandenburg:2005he} where more details can be found.

A significant fraction of the NLSP decays will take place outside the
detector and will thus escape detection.  For the charged slepton NLSP
scenario, however, two recent works have proposed how such NLSPs could
be stopped and collected for an analysis of their decays.  It was
found that up to $\Order(10^3$--$10^4)$ and $\Order(10^3$--$10^5)$ of
charged NLSPs can be trapped per year at the Large Hadron Collider
(LHC) and the International Linear Collider (ILC), respectively, by
placing 1--10~kt of massive additional material around planned
collider detectors~\cite{Hamaguchi:2004df,Feng:2004yi}.

To be specific, we focus on the case where the pure `right-handed'
stau $\stau_{\mathrm R}$ is the NLSP. We assume for simplicity that
the neutralino-stau coupling is dominated by the bino coupling and
that the lightest neutralino is a pure bino.

\subsection{Probing the Peccei--Quinn Scale and the Axino Mass}

\begin{figure*}
\includegraphics[width=9.cm]{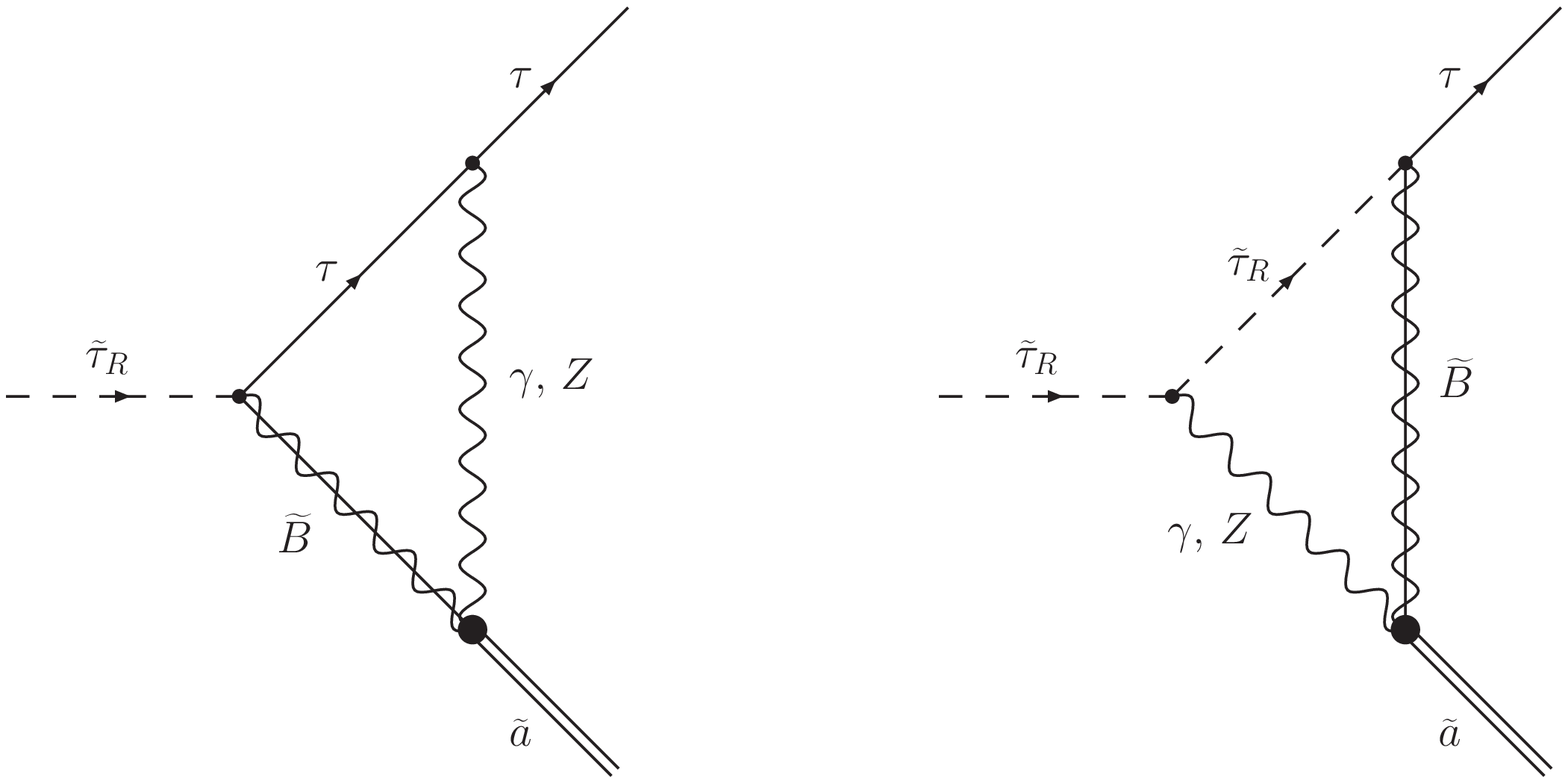}
\caption{
  The 2-body decay $\stau_{\mathrm R}
  \to\tau+\axino$.\label{Fig:2-Body_Decay}}
\end{figure*}

In the axino LSP scenario, the total decay rate of the stau NLSP is
dominated by the 2-body decay $\stau\to\tau+\axino$. In
Fig.~\ref{Fig:2-Body_Decay} we show the corresponding Feynman diagrams
for the considered hadronic (KSVZ) axion models. The heavy KSVZ
(s)quark loops are indicated as effective vertices by thick dots.
The decay rate was estimated as~\cite{Brandenburg:2005he}
\be\label{Eq:Axino2Body}
  \Gamma(\stau_{\mathrm R} \to\tau\,\axino)
  \simeq
  \xi^2\,(25~\mathrm{s})^{-1}
  C_{\rm aYY}^2
  \left(1-\frac{m_{\ax}^2}{m_{\st}^2}\right)
  \left(\frac{m_{\st}}{100\,\GeV}\right)
  \left(\frac{10^{11}\,\GeV}{f_a}\right)^2
  \left(\frac{m_{\tilde{B}}}{100\,\GeV}\right)^2
  \ ,
\ee
where $m_{\tilde{B}}$ is the mass of the bino and $m_{\st}$ is the
mass of the stau NLSP, i.e.\ $m_{\ax} < m_{\st} < m_{\tilde{B}}$.  The
KSVZ-model dependence is expressed by $C_{\rm aYY}\simeq
\mathcal{O}(1)$ and the uncertainty of the estimate is absorbed into
$\xi\simeq\Order(1)$.  Thus, from the lifetime of the stau NLSP,
$\tau_{\st} \approx 1/\Gamma(\stau_{\mathrm R} \to\tau\,\axino)$, an
estimate of the Peccei--Quinn scale $f_a$ can be
obtained~\cite{Brandenburg:2005he}
\be\label{Eq:PQ_Scale}
  f_a^2 
  \simeq
  \left(\frac{\tau_{\st}}{25~\mathrm{s}}\right)\, 
  {\xi^2\,C_{\rm aYY}^2} 
  \left(1-\frac{{m_{\ax}^2}}{{m_{\st}^2}}\right)
  \left(\frac{m_{\st}}{100\,\GeV}\right)
  \left(\frac{m_{\Bi}}{100\,\GeV}\right)^2
  \left(10^{11}\,\GeV\right)^2
  \ .
\ee
Indeed, we expect that $m_{\st}$ and $m_{\Bi}$ will already be known
from other processes when the stau NLSP decays are analyzed. The
dependence on $m_{\ax}$ is negligible for $m_{\ax}/m_{\st}\lesssim
0.1$.  For larger values of $m_{\ax}$, the stau NLSP decays can be
used to determine the axino mass kinematically, i.e., from a
measurement of the energy of the emitted tau $E_\tau$,
\be\label{Eq:Axino_Mass} 
  m_{\ax} =
  \sqrt{{m_{\st}^2}+{m_\tau^2}-2{m_{\st} E_\tau}} 
  \ ,
\ee
where the error is given by the experimental uncertainty on $m_{\st}$
and $E_\tau$. The determination of both the Peccei--Quinn scale $f_a$
and the axino mass $m_{\ax}$ is crucial for insights into the
cosmological relevance of the axino LSP.

\subsection{Measuring the Gravitino Mass and the Planck Scale}

In the gravitino LSP scenario, the main decay mode of the stau NLSP is
the 2-body decay $\stau\to\tau+\gravitino$. Neglecting the $\tau$
mass, the following tree-level result for the decay rate is obtained
from the SUGRA Lagrangian
\be\label{Eq:Gravitino2Body}
 \Gamma(\stau_{\mathrm R}\to\tau\,\gravitino)
  =
  \frac{m_{\st}^5}{48\pi\,m_{\gr}^2\,\MPl^2}
  \left(
  1 - \frac{m_{\gr}^2}{m_{\st}^2}
  \right)^4
  =
  (5.89~\mathrm{s})^{-1}
  \left(\frac{m_{\st}}{100~\mathrm{GeV}}\right)^5
  \left(\frac{10~\mathrm{MeV}}{m_{\gr}}\right)^2
  \left(
  1 - \frac{m_{\gr}^2}{m_{\st}^2}
  \right)^4
\ee
with the value of the reduced Planck mass
$\MPl = (8\pi\,G_{\rm N})^{-1/2} = 2.435\times 10^{18}\,\GeV$ 
as given by macroscopic measurements of Newton's
constant~\cite{Eidelman:2004wy}
$G_{\rm N} = 6.709\times 10^{-39}\,\GeV^{-2}$.
Thus, the gravitino mass $\mgravitino$ can be determined once the stau
NLSP lifetime, $\tau_{\st} \approx 1/\Gamma(\stau_{\mathrm R}
\to\tau\,\gravitino)$, and $m_{\st}$ are measured. This will be
crucial for insights into the SUSY breaking mechanism. If the
gravitino mass can be determined independently from the kinematics via
$m_{\gr} = ({m_{\st}^2}+{m_\tau^2}-2{m_{\st} E_\tau})^{1/2}$,
the lifetime $\tau_{\st}$ can also be used for a microscopic
measurement of the Planck scale~\cite{Buchmuller:2004rq}
\be\label{Eq:Planck_Scale}
  \MPl^2 =  
  \frac{\tau_{\st}}{48\pi} 
  \frac{m_{\st}^5}{m_{\gr}^2}
  \left(
  1 - \frac{m_{\gr}^2}{m_{\st}^2}
  \right)^4
  \ .
\ee
If consistent with macroscopic measurements, this would provide
evidence for the existence of SUGRA in nature.

\subsection{Distinguishing between Axinos and Gravitinos}

A question arises as to whether one can distinguish between the axino
LSP and the gravitino LSP scenarios.  For $m_{\st}=100\,\GeV$ and
$m_{\Bi}=110\,\GeV$, for example, the stau NLSP lifetime in the axino
LSP scenario can range from $\Order(0.01~{\mbox{s}})$ for $f_a=5\times
10^9\,\GeV$ to $\Order(10~{\mbox{h}})$ for $f_a=5\times
10^{12}\,\GeV$. In the gravitino LSP case, the corresponding lifetime
can vary over an even wider range, e.g., from $6\times 10^{-8}\,{\rm
  s}$ for $\mgravitino = 1~\keV$ to 15~years for $\mgravitino =
50~\GeV$. Thus, both a very short lifetime, $\tau_{\st} \lesssim$~ms,
and a very long one, $\tau_{\st} \gtrsim$~days, will point to the
gravitino LSP. On the other hand, if the LSP mass cannot be measured
and the lifetime of the stau NLSP is within the range
$\Order(0.01~{\mbox{s}})$--$\Order(10~{\mbox{h}})$, it will be very
difficult to distinguish between the axino LSP and the gravitino LSP
from the lifetime $\tau_{\st}$ alone.

\begin{figure*}
\fbox{\includegraphics[width=85.mm]{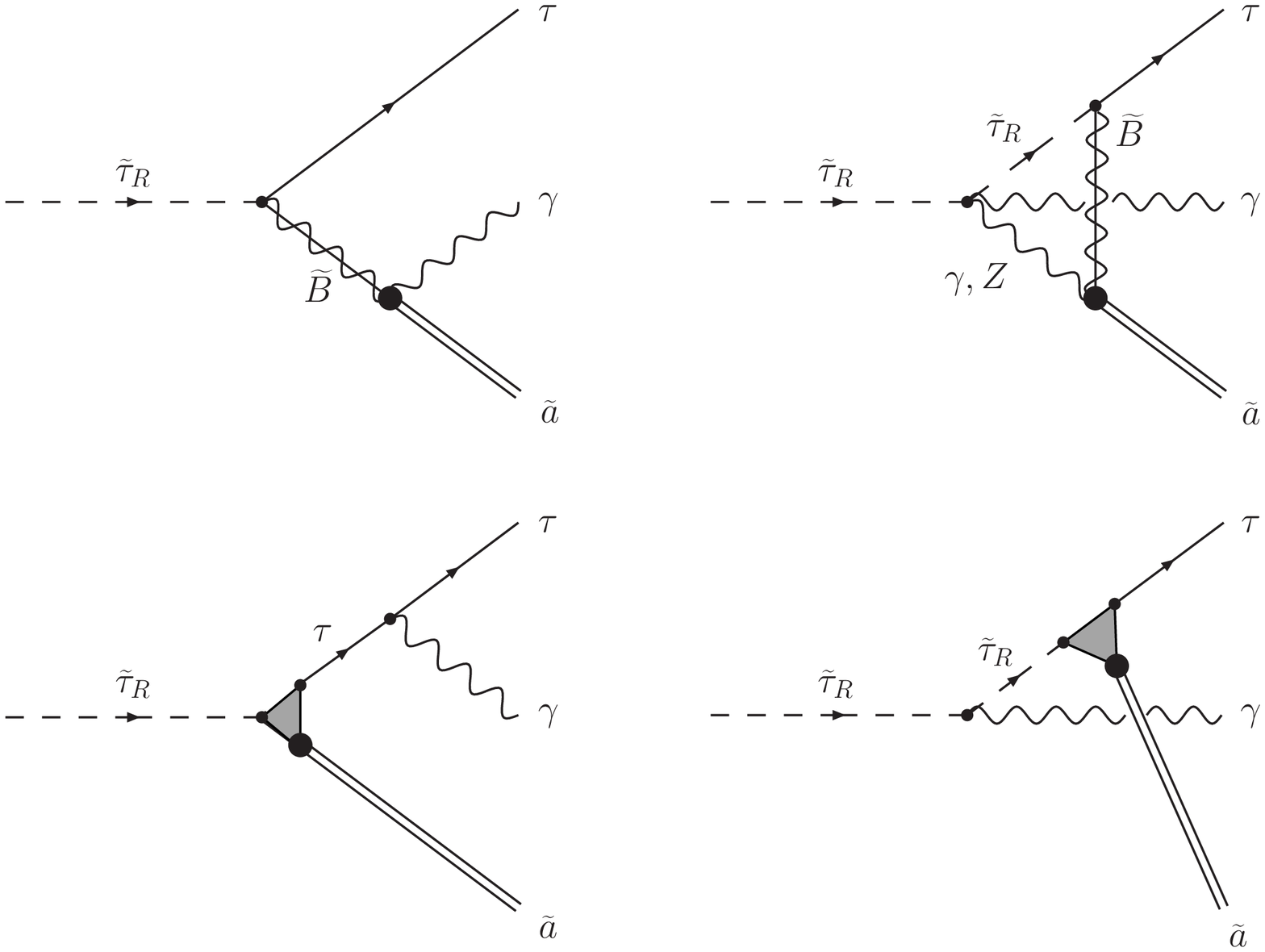}}
\hskip 0.2cm
\fbox{\includegraphics[width=85.mm]{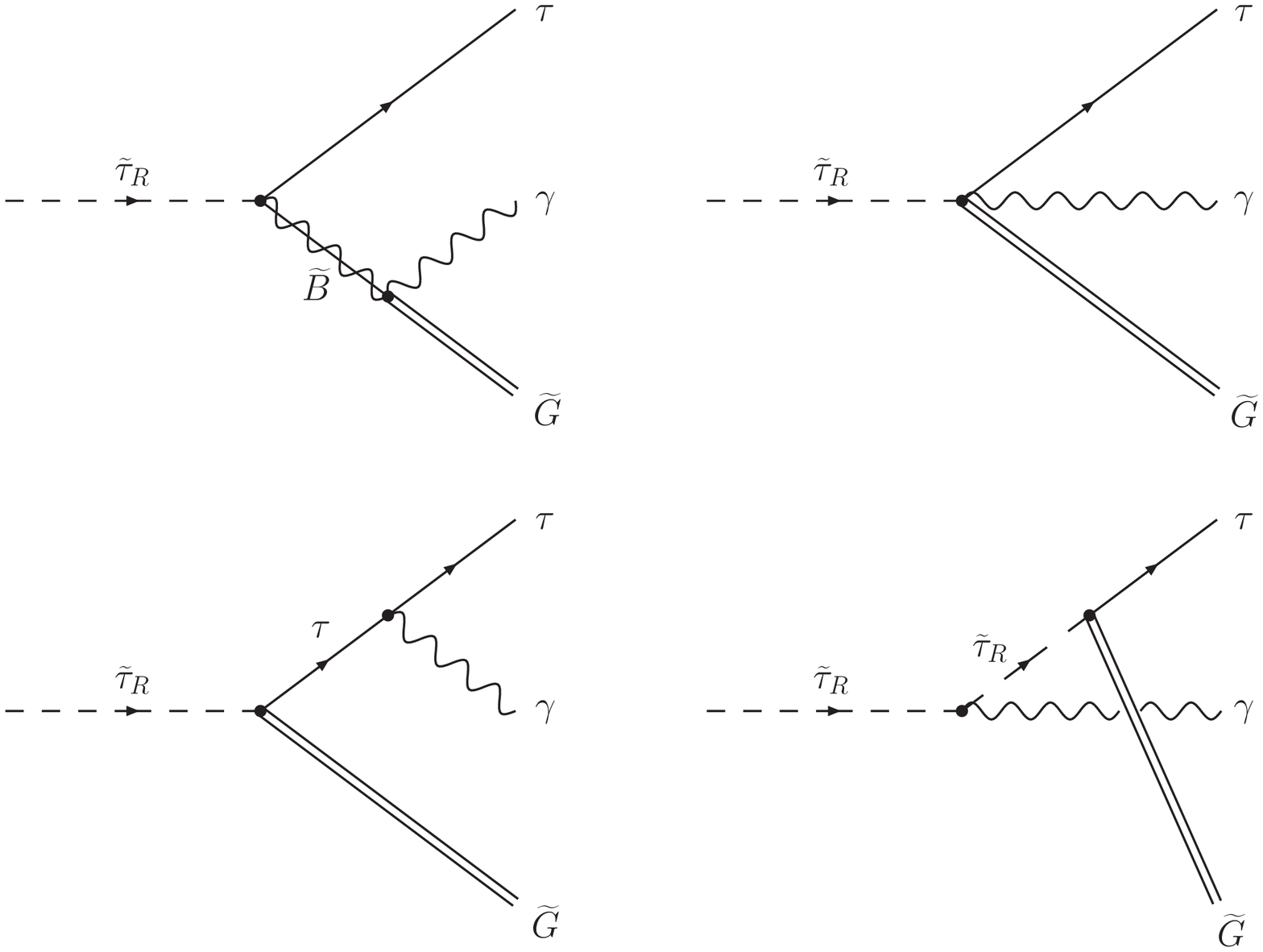}}
\caption{
  The 3-body decays $\stau_{\mathrm R} \to \tau + \gamma + \axino$
  (left) and $\stau_{\mathrm R} \to \tau + \gamma + \gravitino$
  (right).\label{Fig:3-Body_Decays}}
\end{figure*}

\begin{figure*}
\includegraphics[width=7.cm]{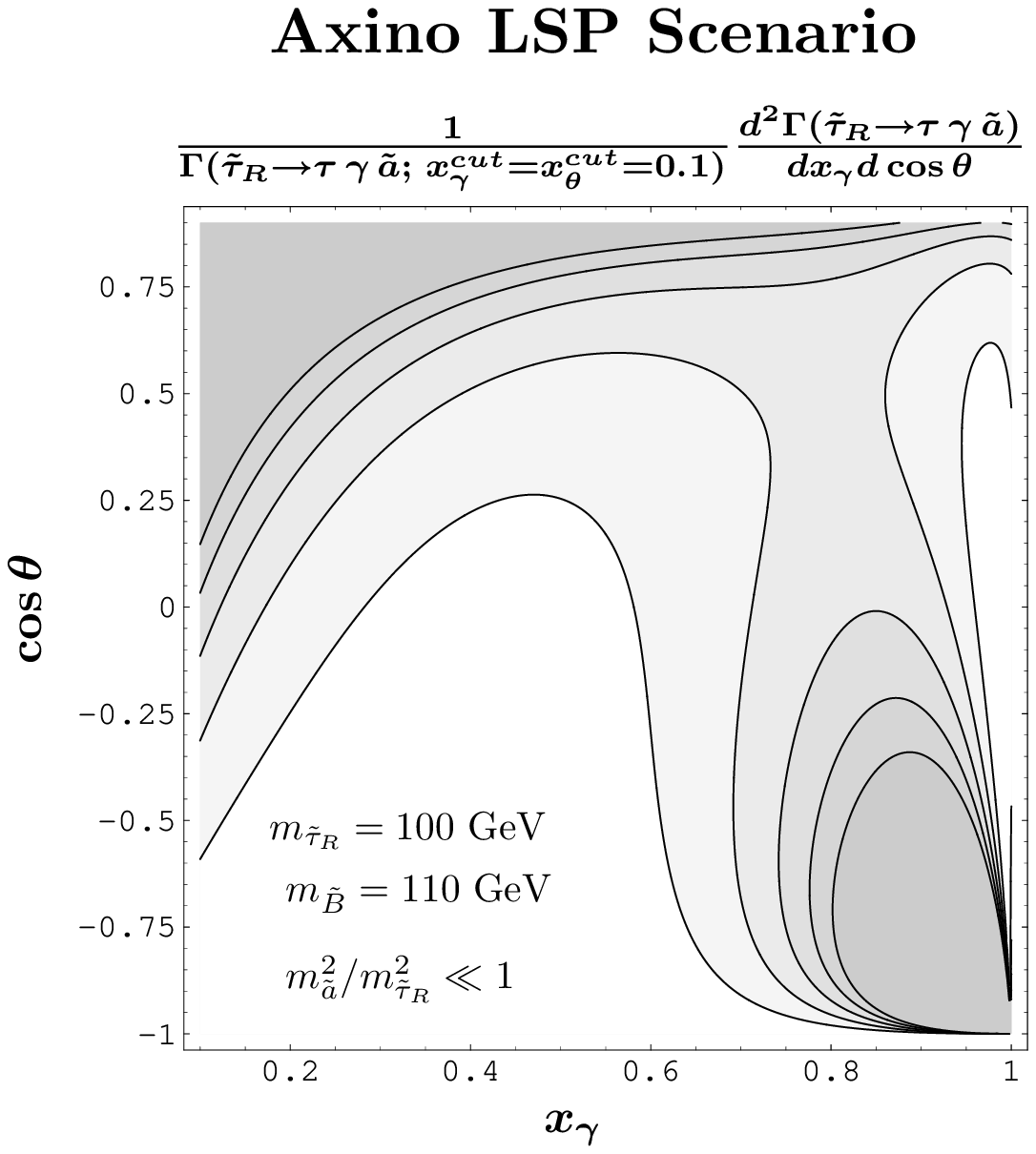}
\hskip 1.5cm
\includegraphics[width=7.cm]{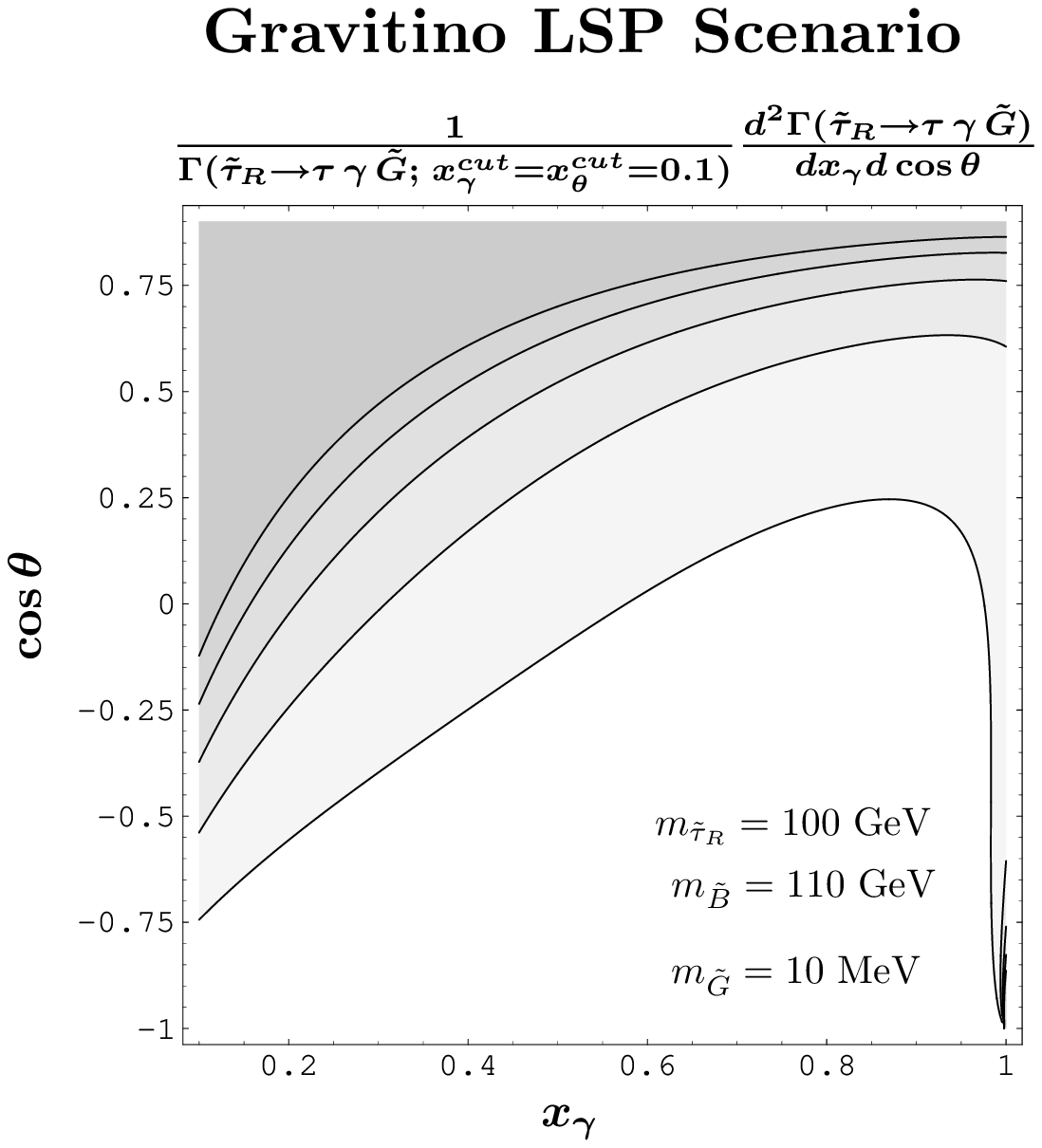}
\caption{
  Differential distributions of the visible decay products in
  $\stau_{\mathrm R} \to \tau + \gamma + \axino$ (left) and
  $\stau_{\mathrm R} \to \tau + \gamma + \gravitino$
  (right).\label{Fig:Diff_Distributions}}
\end{figure*}

The situation is considerably improved when one considers the 3-body
decays $\stau_{\mathrm R} \to \tau + \gamma + \axino/\gravitino$. The
Feynman diagrams of the dominant contributions are shown in
Fig.~\ref{Fig:3-Body_Decays}, where thick dots represent heavy KSVZ
(s)quark loops and shaded triangles the set of diagrams given in
Fig.~\ref{Fig:2-Body_Decay}.
From the corresponding differential rates~\cite{Brandenburg:2005he},
one obtains the differential distributions of the visible decay
products. These are illustrated in Fig.~\ref{Fig:Diff_Distributions}
in terms of the quantity
\be\label{Eq:Fingerprint}
        {1 
        \over 
        \Gamma(\stau_{\mathrm R}\to\tau\,\gamma\, i\,;
        x_{\gamma}^{\mathrm{cut}},x_{\theta}^{\mathrm{cut}})}
        \,\,
        {d^2\Gamma(\stau_{\mathrm R}\to\tau\,\gamma\, i)
        \over
        d x_{\gamma}d \cos\theta}
        \ ,
        \quad 
        i = \axino, \gravitino
        \ ,
\ee
where $x_\gamma\equiv 2 E_\gamma/m_{\st}$ is the scaled photon energy,
$\theta$ is the opening angle between the photon and tau directions,
and
\be
\Gamma(\stau_{\mathrm R}\to\tau\,\gamma\,i\,;
x_\gamma^{\mathrm{cut}},x_\theta^{\mathrm{cut}})
  \equiv
  \int^{1-A_{i}}_{x_\gamma^{\mathrm{cut}}}
  d x_\gamma
  \int^{1-x_\theta^{\mathrm{cut}}}_{-1}
  d \cos\theta \,\,
  \frac{d^2\Gamma(\stau_{\mathrm R}\to\tau\,\gamma\,i)}{dx_\gamma d\cos\theta}
        \ 
        \quad \mbox{with} \quad
        A_{i} \equiv \frac{m_{i}^2}{m_{\st}^2}
\ee
is the respective integrated 3-body decay rate with the cuts $x_\gamma
> x_\gamma^{\mathrm{cut}}$ and $\cos\theta <
1-x_\theta^{\mathrm{cut}}$.
Note that the quantity~(\ref{Eq:Fingerprint}) is independent of the
2-body decay, the total NLSP decay rate, and the Peccei--Quinn/Planck
scale.
%

The figure shows the normalized differential
distributions~(\ref{Eq:Fingerprint}) for the axino LSP with
$m_{\ax}^2/m_{\stau}^2 \ll 1$ (left) and the gravitino LSP with
$\mgravitino = 10\,\MeV$ (right), where $m_{\st} = 100\,\GeV$,
$m_{\Bi} = 110\,\GeV$, and $x_\gamma^{\mathrm{cut}} =
x_\theta^{\mathrm{cut}}=0.1$. The contour lines represent the values
0.2, 0.4, 0.6, 0.8, and 1.0, where the darker shading implies a higher
number of events.
In the case of the gravitino LSP, the events are peaked only in the
region where the photons are soft and the photon and the tau are
emitted with a small opening angle ($\theta\simeq 0$).  In contrast,
in the axino LSP scenario, the events are also peaked in the region
where the photon energy is large and the photon and the tau are
emitted back-to-back ($\theta \simeq \pi$).
Thus, if the observed number of events peaks in both regions, there is
strong evidence for the axino LSP and against the gravitino LSP.

To be specific, with $10^4$ analyzed stau NLSP decays, we expect about
165$\pm$13 (stat.) events for the axino LSP and about 100$\pm$10
(stat.) events for the gravitino LSP~\cite{Brandenburg:2005he}, which
will be distributed over the corresponding
($x_\gamma$,\,$\cos\theta$)-planes shown in
Fig.~\ref{Fig:Diff_Distributions}. In particular, in the region of
$x_{\gamma}\gtrsim 0.8$ and $\cos\theta \lesssim -0.3$, we expect
about 28\% of the 165$\pm$13 (stat.) events in the axino LSP case and
about 1\% of the 100$\pm$10 (stat.) events in the gravitino LSP case.
These numbers illustrate that $\Order(10^4)$ of analyzed stau NLSP
decays could be sufficient for the distinction based on the
differential distributions. To establish the feasibility of this
distinction, a dedicated study taking into account the details of the
detectors and the additional massive material will be crucial, which
we leave for future studies.

Some comments are in order. 
The differences between the two scenarios shown in
Fig.~\ref{Fig:Diff_Distributions} become smaller for larger values of
$m_{\Bi} / m_{\st}$.  This ratio, however, remains close to unity for
the stau NLSP in unified models.
Furthermore, if $m_{\gr} < m_{\ax} < m_{\st}$ and $\Gamma(\stau
\to \axino\,X) \gg \Gamma(\stau \to \gravitino\,X)$, one would still
find the distribution shown in the left panel of
Fig.~\ref{Fig:Diff_Distributions}.
The axino would then eventually decay into the gravitino LSP and the
axion.  Conversely, the distribution shown in the right panel of
Fig.~\ref{Fig:Diff_Distributions} would be obtained if $m_{\ax} <
m_{\gr} < m_{\st}$ and $\Gamma(\stau \to \axino\,X) \ll
\Gamma(\stau \to \gravitino\,X)$. Then it would be the gravitino that
would eventually decay into the axino LSP and the axion.
Barring these caveats, the signatures shown in
Fig.~\ref{Fig:Diff_Distributions} will provide a clear distinction
between the axino LSP and the gravitino LSP scenarios.

\section{CONCLUSIONS}

Axino/gravitino LSPs from thermal production in the early Universe can
provide the right amount of cold dark matter depending on their mass
and the reheating temperature after inflation. If a charged slepton is
the NLSP, future colliders can provide signatures of axino/gravitino
LSPs and other insights into physics beyond the MSSM. 

\begin{acknowledgments}
  
  I am grateful to A.~Brandenburg, L.~Covi, K.~Hamaguchi, and
  L.~Roskowski for an enjoyable collaboration.

\end{acknowledgments}


\end{document}